## Summary

Comparing the obtained experimental data for the proton magnetic relaxation with the microcalorimetric ones it is concluded that in the aqueous solution of collagen + DNA at temperatures below 35°C a stable molecular complex collagen-DNA is formed prohibiting the formation of the hydration cover of collagen which is manifested by the almost two times decrease of hydration.

At temperatures (25-45)°C the disruption of complex and melting of triple helix of collagen takes place. Starting from temperature 45°C up to 90°C hydration values $\omega_{ex}$, calculated from experimental data on proton relaxation for solution and $\omega_T$, theoretical values calculated for the solution coincide speaking of the fact that macromolecules in the solution are not practically interacting with each other and not influencing the creation of each others hydration covers.

**G.M. Mrevlishvili[1], D.V. Svintradze[1], N.O. Dzhariashvili[1], L.O. Namisheishvili[1], G.I. Mamniashvili[2], Yu.G. Sharimanov[2], T.O. Gegechkori[2]**

[1] Department of Physics, Tbilisi State University, Chavchavadze av.3, Tbilisi 0128, Georgia
[2] E. Andronikashvili Institute of Physics, Tamarashvili str. 6, Tbilisi 0177, Georgia

## Collagen-DNA macromolecular complex study by the proton magnetic relaxation method.

The proton magnetic relaxation method was used to study diluted aqueous DNA and collagen solutions in a number of works [1-3], where it was obtained exhaustive information on the hydration of investigated macromolecules and mobility of water molecules belonging to different categories of solvents. One bears in mind so-called "rigidly"-nonrotationally bounded water, "bounded" water and "free" water. In particular, the use of proton magnetic relaxation method in combination with the microcalorimetry [3] made it possible to study in details the conformational changes and the structure of the hydration shell of collagen and to present the three-fractional hydration model of given protein's macromolecules.

In work [4] using the microcalorimetry method it was measured the thermodynamic parameters of the complex between triple helix of collagen and double helix of DNA in the diluted aqueous solution. It was shown that interaction between collagen and DNA induces the destruction of hydration shell of the collagen triple helix and stabilization of ds-DNA including its hydration shell.

During the formation of complex it appears the cluster structure of water stabilizing the complex. The complex which can be considered as a selfassembled system is spontaneously formed in the aqueous solution.

It was shown [5] that the formation of this complex is caused by the interaction between negatively charged $PO_4^{-2}$ phosphate groups arranged on ds-DNA and dipole moments of amino-imino acids, and the interaction with positively charged lysine residue.

The twisted fiber structure of collagen-DNA complex is connected with the hydrogen bonds between CH acceptor groups of the collagen triple helix and $PO_4$ donor groups of DNA, as well as with the hydrophobic interactions [5].

The complex is characterized by a high degree of arrangement and is a fiber like structure where DNA with a help of phosphate gropes arranged on ds-DNA plays the role of a polymer matrix for collagen fibrils [5].

In present work, with the aim to specify the structure characteristics of revealed complex, it is carried out the investigation of proton magnetic relaxation in the range of melting temperatures of triple helix of collagen and double helix of DNA in the diluted aqueous solution of mixture of the above pointed biopolymers.

NMR measurements were performed using a Minispec p20 pulse relaxometer (Bruker) at a resonance frequency of 19,8 MHz. Time for spin-lattice relaxation for protons $T_1$ was measured using pulse sequence $180°$ -$\tau$ - $90°$ [6]. The spin-spin relaxation time $T_2$ for protons was determined using the method of Karr – Purcell – Meilboom – Gill (KPMG) [7]. Experimental errors in relaxation time were less then 5%. Investigated solutions of biopolymers, rat skin collagen and DNA calf thymus were prepared using a solvent of the following condition: 0.015 citric buffer, 0.15 M NaCl, pH=4.0.

In fig.1 it was presented the results of $T_1$, $T_2$ times measurements in the aqueous solution of collagen on temperature in the range from $20°C$ to $90°C$. As it is seen from 1, 2 curves in the temperature range $(34 – 37)°C$ it is observed a sharp increase of relaxation times on the monotonic temperature rise background of their values. This sharp increase accordingly [4] is also accompanied by a thermoabsorption pick and a jump of partial heat capacity, which in its turn is caused by the intramolecular conformation transition "triple helix – random coil" [8].

Further in fig. 2, it was shown the temperature dependences $T_1$, $T_2$ for DNA aqueous solution in the same temperature interval. At the temperatures of the order of $74°C$ in the investigated solution proton relaxation times are also sharply increased (curves 1, 2). Besides this, on the corresponding calorimetric curve [4] it is observed the thermoabsorption pick which is not accompanied by a jump of the partial heat capacity. As it was established in works [2,4], such behavior of relaxation times and thermodynamic parameters of the investigated solution is connected with a conformation transition of DNA molecules from the state of double helix to the state of random coil [3].

In works [1-3] for the interpretation of proton magnetic relaxation data in the aqueous solutions of biopolymers the two-component model of hydration was used which in the definite cases makes it possible, using experimentally measured $T_1$, $T_2$ for the solution and solvent separately, to determine the hydration parameters of biomolecules at different temperatures of investigated systems. The procedure of calculation is described in details in work [1].

Using the above mentioned calculation method and our experimental data (fig. 1, fig. 2) we obtained the hydration values $\omega_C$ and $\omega_D$, for collagen and DNA, correspondingly, in grams on biopolymer's gram at the definite having a particular meaning temperatures (see table 1).

As it is seen from the table, at temperature $35°C$ for collagen and at temperature $74°C$ for DNA it takes place a step-like decrease of hydrations $\omega_C$ and $\omega_D$, accompanying the disruption of triple and double helices and transition in the coiled state. The obtained data are in a good quantitative agreement with the data of works [2, 3] at the corresponding concentrations of biopolymers in investigated solutions.

It is obvious that starting from the obtained values for $\omega_C$ and $\omega_D$ one could theoretically calculated anticipated hydration $\omega_T$ for an aqueous solution non-interacting with each other collagen and DNA molecules accordingly formulae:

$$\omega_T = \frac{\omega_C C_C + \omega_D C_o}{C_C + C_o}, \qquad (1)$$

where $C_C$ and $C_o$ are concentrations in g/g $H_2O$ of collagen and DNA, correspondingly.

Anticipated values of hydration $\omega_T$, calculated by the expression (1), for the solution 0,45 mg collagen/ml $H_2O$ + 2 mg DNA/ml $H_2O$ for different temperature are presented in table 1.

In fig.3 it is shown measured by us temperature dependences of $T_1$, $T_2$ times for the aqueous solution of collagen and DNA in the above pointed concentration relation (curves 1, 2). On the experimental curves 2, 3 (fig.3) in the temperature range $(35 - 45)^\circ$ C, it is seen the clear break which corresponds by its temperature location to the thermoabsoption pick at temperature $40^\circ$C on the microcalorimetric curve in [4]. Further at temperatures near $74^\circ$C, it is seen a jump of relaxation times $T_1$, $T_2$ also coinciding by its location with the thermoabsorption pick corresponding to DNA melting [4].

The measured dependences make it possible using method [1] also in this case to calculate $\omega_{ex}$ - the real averaged hydration of biomacromolecules in the aqueous solution collagen + DNA. Results of these calculations are presented in table 1. Comparing obtained from the proton relaxation experiment values of $\omega_{ex}$ with the calculated ones for the solution of noninteracting macromolecules $\omega_T$, one could note that at temperatures $(30 - 35)^\circ$C hydration of real solution collagen + DNA almost in 2 times lower then the anticipated hydration $\omega_T$ and is approximately equal to the hydration of DNA molecules. At further increase of temperature up to $90^\circ$C values $\omega_{ex}$ and $\omega_T$ are practically coinciding.

Summing up above presented facts and comparing data for the proton magnetic relaxation with microcalorimetric ones [4], we arrive to a conclusion that in the aqueous solution of collagen + DNA at temperatures below $35^\circ$C a stable molecular complex collagen - DNA is formed prohibiting the formation of hydration cover of collagen which is manifested by the almost two times decrease of hydration. At temperatures $(35 - 45)^\circ$C the disruption of complex and melting of the triple helix of collagen takes place. Starting from temperature $45^\circ$C up to $90^\circ$C, hydration values $\omega_{ex}$ and $\omega_T$ are practically coincide what speaks of the fact that macromolecules in the solution are not practically interacting with each other and not influencing the creation of each others hydration covers.

**Table. 1.** Hydration of macromolecules at different temperatures.

$\omega_c$ - hydration of collagen accordingly proton relaxation data (fig.1).

$\omega_D$ - hydration of DNA accordingly proton relaxation data (fig.2).

$\omega_T$ - theoretically calculated hydration for the solution of non-interacting molecules
    (0,45 mg collagen/ml $H_2O$ + 2 mg DNA/ml $H_2O$).

$\omega_{ex}$ - hydration calculated from experimental data on the proton relaxation for solution
    (0,45 mg collagen/ml $H_2O$ + 2 mg DNA/ml $H_2O$) (fig.3).

| Temperature t °C | $\omega_C$ gH$_2$O/g collagen | $\omega_D$ gH$_2$O/g DNA | $\omega_T$ gH$_2$O/g Biopolymer | $\omega_{ex}$ gH$_2$O/g Biopolymer |
|---|---|---|---|---|
| 30 | 0,65 | 0,142 | 0,235 | 0,131 |
| 35 | 0,636 | 0,114 | 0,209 | 0,114 |
| 45 | 0,33 | 0,104 | 0,146 | 0,145 |
| 60 | 0,31 | 0,097 | 0,135 | 0,137 |
| 65 | 0,32 | 0,093 | 0,135 | 0,134 |
| 75 | 0,33 | 0,07 | 0,066 | 0,066 |
| 90 | 0,34 | 0,05 | 0,066 | 0,066 |

FIGURE CAPTIONS

Fig. 1. Temperature dependences of spin-lattice ($T_1$) (curve 1) and spin-spin ($T_2$) (curve 2) relaxation times of protons in aqueous collagen solution (0,9 mg/ml) in the range of triple helix melting.

Fig. 2. Temperature dependences of spin-lattice ($T_1$) (curve 1) and spin-spin ($T_2$) (curve 2) relaxation times of protons in aqueous DNA solution (4 mg/ml) in the range of double helix melting.

Fig.3. Temperature dependences of spin-lattice ($T_1$) (curve 1) and spin-spin ($T_2$) (curve 2) relaxation times of protons in aqueous solution of biopolymers mixture with concentration (0,45 mg/ml for collagen and 2 mg/ml for DNA) in the temperature range of existence and disruption of collagen – DNA complex [4] .

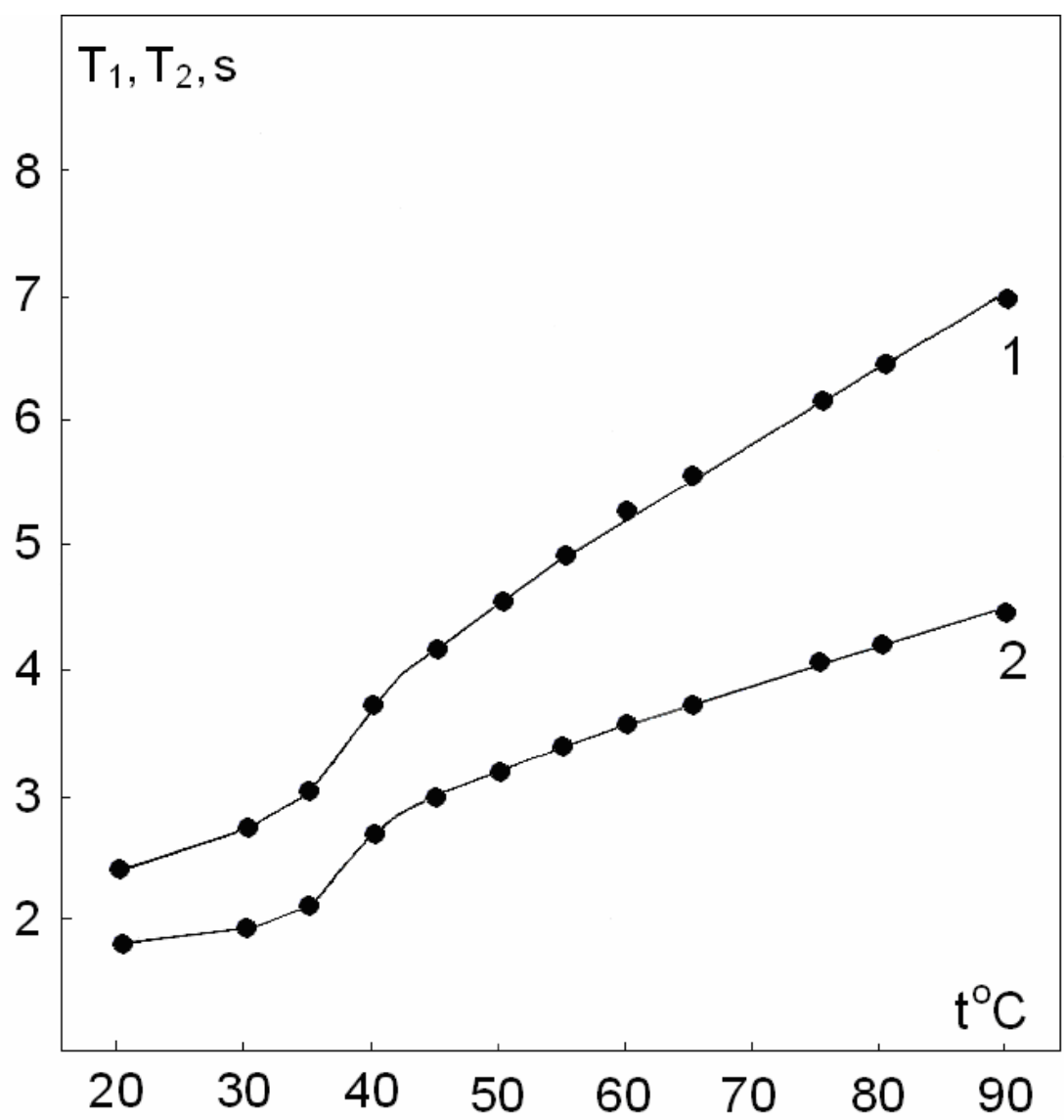

Fig.1

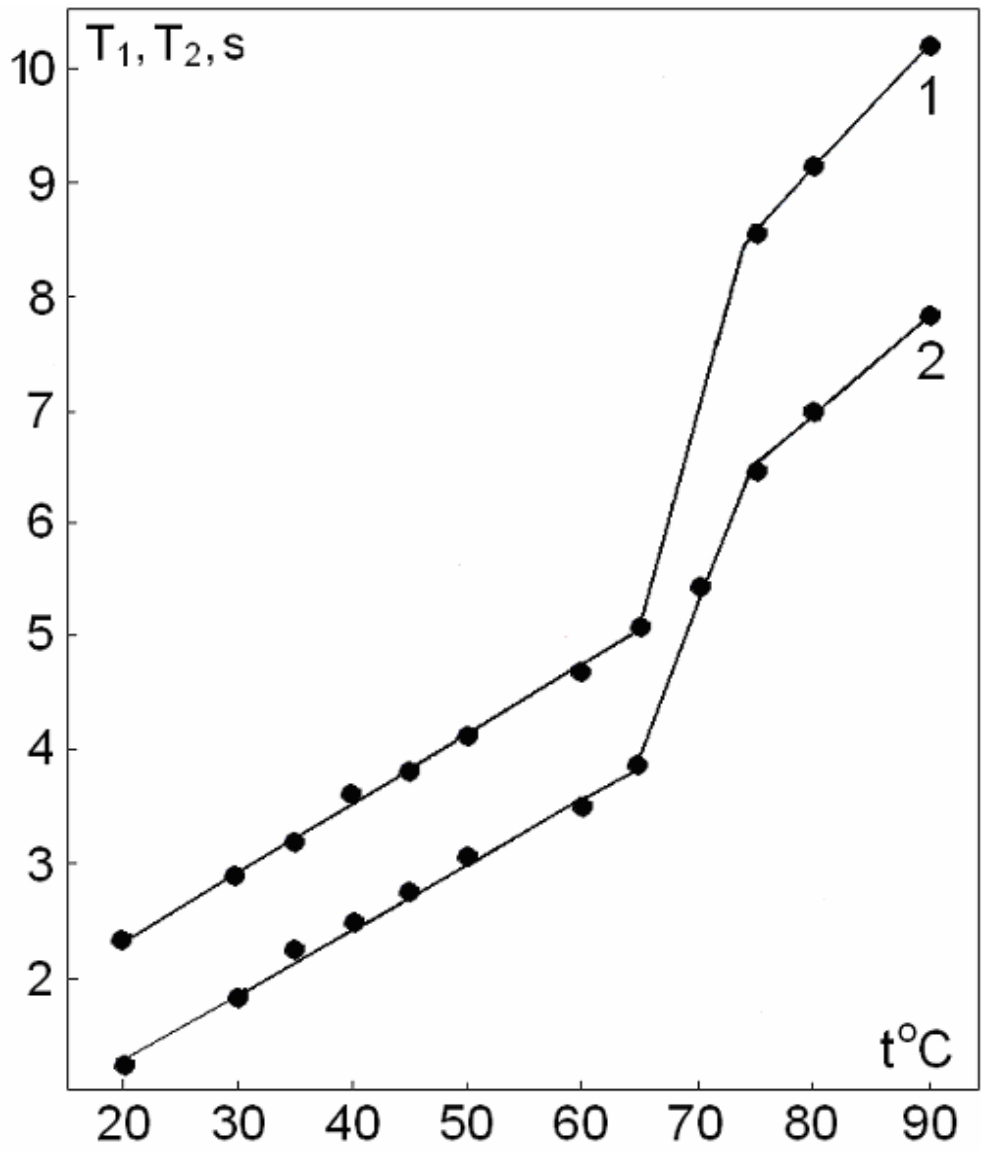

Fig.2

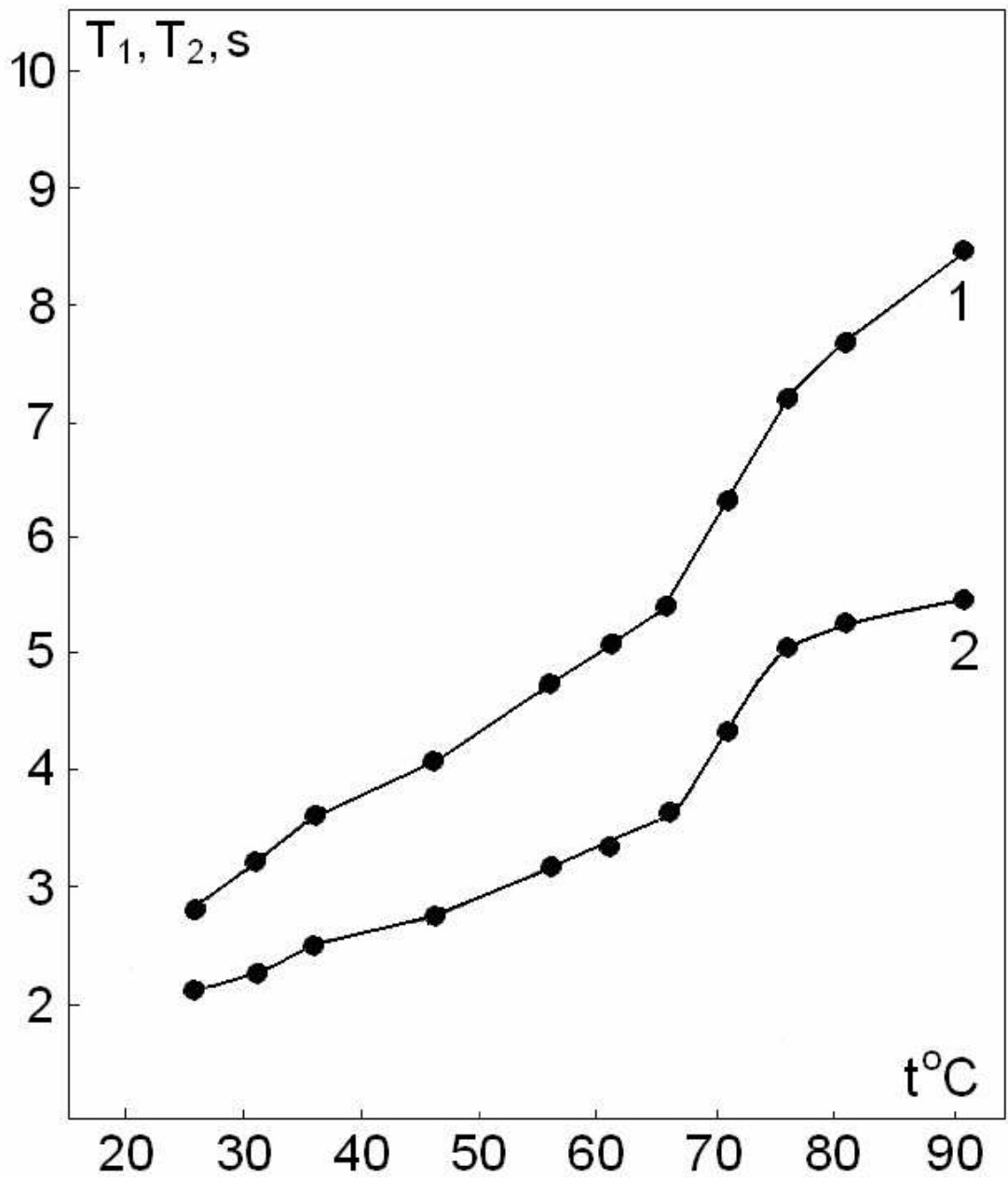

Fig.3